# COLD REACTION VALLEYS IN THE RADIOACTIVE DECAY OF SUPERHEAVY $^{286}112$, $^{292}114$ AND $^{296}116$ NUCLEI

**K. P. Santhosh, S. Sabina**

*School of Pure and Applied Physics, Kannur University, Payyanur Campus, India*

Cold reaction valleys in the radioactive decay of superheavy nuclei $^{286}112$, $^{292}114$ and $^{296}116$ are studied taking Coulomb and Proximity Potential as the interacting barrier. It is found that in addition to alpha particle, $^{8}$Be, $^{14}$C, $^{28}$Mg, $^{34}$Si, $^{50}$Ca, etc. are optimal cases of cluster radioactivity since they lie in the cold valleys. Two other regions of deep minima centered on $^{208}$Pb and $^{132}$Sn are also found. Within our Coulomb and Proximity Potential Model half-life times and other characteristics such as barrier penetrability, decay constant for clusters ranging from alpha particle to $^{68}$Ni are calculated. The computed alpha half-lives match with the values calculated using Viola--Seaborg--Sobiczewski systematics. The clusters $^{8}$Be and $^{14}$C are found to be most probable for emission with $T_{1/2} < 10^{30}$s. The alpha-decay chains of the three superheavy nuclei are also studied. The computed alpha decay half-lives are compared with the values predicted by Generalized Liquid Drop Model and they are found to match reasonably well.

# 1. INTRODUCTION

Generally radioactive nuclei decay through alpha and beta decay with subsequent emission of gamma rays in many cases. Again since 1939 it is well known that many radioactive nuclei also decay through spontaneous fission. In 1980 a new type of decay known as cluster radioactivity was predicted by Sandulescu et al. [1] on the basis of Quantum Mechanical Fragmentation Theory (QMFT) [2, 3]. Cluster radioactivity is the spontaneous decay of nuclei by the emission of particles heavier than alpha particle say $^{14}$C, $^{24}$Ne, $^{30}$Mg and $^{34}$Si and therefore occupies intermediate position between alpha decay and spontaneous fission. The first experimental identification of cluster decay was accomplished by Rose and Jones [4] in 1984 in the radioactive decay of $^{223}$Ra by $^{14}$C emission with a half-life of 3.7 ± 1.1 years. Since then intense experimental research has led to the detection of about 20 cases of spontaneous emission of clusters ranging from $^{14}$C to $^{34}$Si from trans-lead nuclei with partial half-life ranging from $10^{11}$s to $10^{28}$s. The common feature of all these emissions is that heavier nuclei emit heavier fragments in such a way that the daughter nuclei are always the doubly-magic $^{208}$Pb or closely neighbouring nuclei. In 1991 a new island of proton-rich parent nuclei with $Z$ = 56--64 and $N$ = 56--72 exhibiting cluster decay leading to daughter nuclei close to the doubly magic $^{100}$Sn has been predicted by Poenaru et al. [5] on the basis of Analytical Super Asymmetric Fission Model (ASAFM). This was later confirmed experimentally by Oganessian et al. [6, 7] at Dubna and by Guglielmetti et al. [8, 9] at GSI, Darmstat. Thus cluster decay is now established as one of the key decay modes for radioactive nuclei.

The study of superheavy elements (SHE) leads to many new findings, the most important among them being the possible appearance of new magic shell numbers or more precisely the prediction of the doubly-magic nucleus next to $Z = 82$, $N = 126$, $^{208}$Pb. A number of studies have been done to this effect leading to the prediction of new magic numbers in the superheavy region. The first papers that predicted the doubly magic nucleus with $Z = 114$ and $N = 184$ were that of Sobiczewski et al. [10] and Meldner [11]. Later in 1969, Nilsson et al. [12] calculated the nuclear potential energy surfaces as a function of deformations on the basis of a modified oscillator model. These calculations were extended to the predicted superheavy region around $Z = 114$ and $N = 184$ and the total overall stability with respect to alpha decay, beta decay, and spontaneous fission was found to be most favorable in the vicinity of $Z = 110$ and $N = 184$. Concurrently Mosel et al. [13] also predicted islands of stability around $Z = 114$ and $N = 184$ in the superheavy region. Using the Skyme--Hartree--Fork method Cwiok et al. [14] in 1996, investigated the ground state properties of super heavy nuclei with $108 \leq Z \leq 128$ and $150 \leq N \leq 192$ and the detailed analysis of the shell effects were given. In 1997, Rutz et al. [15] scrutinized the shell structure of superheavy nuclei within the relativistic mean field model and their search was in the region $Z = 110$--140 and $N = 134$--298. The authors found the doubly-magic spherical nuclei at ($Z = 114$, $N = 184$), ($Z = 120$, $N = 172$) or at ($Z = 126$, $N = 184$). Using Deformed Relativistic Mean Field calculation, Patra et al. [16] have predicted $Z = 120$, $N = 184$ as the next possible magic numbers in the superheavy region. Duarte et al. [17] noted a local minimum around $N = 184$ for nuclei with $Z = 110$--121 in their study of the elements in the region of the $ZN$ plane defined by $155 \leq N \leq 220$ and $110 \leq Z \leq 135$ using Effective Liquid Drop Model (ELDM). They attribute this local minimum to a spherical neutron shell closure at $N = 184$. On the basis of the relativistic mean field approach by using the axially deformed harmonic oscillator basis, Sharma et al. [18] predict $Z = 120$, $N = 172$, $^{292}120$ nucleus as doubly magic and $Z = 120$, $N = 184$, $^{304}120$ nucleus as a two cluster configuration. On the basis of Preformed Cluster Model, Gupta et al. [19] found that the α-decay half-life for $Z = 114$, $A = 285$ nucleus is very high which means that $^{285}114$ is very stable against α decay. This stability is attributed either to the magicity of protons at $Z = 114$ or of neutrons at $N = 172$ or to both, perhaps more so to $Z = 114$ since $N = 172$ is predicted to be magic only when $Z = 120$. In the light of these findings cluster radioactivity has a role to play in superheavy element studies because almost all the residual nuclei in cluster decay have been found to be doubly magic. But the exploration of cluster radioactivity in superheavy region did not receive much attention for a long time. One reason for this is attributed to the instability of nuclei in this region. Even in this event, from the theoretical point of view, the extension of the periodic table towards the superheavy island of stability is very important for testing and developing nuclear structure models as one can see in Denisov et al. [20]. Also the half-lives of different radioactive decay modes such as α decay, cluster radioactivity and spontaneous fission are important to identify the decay chains of superheavy elements, which are the experimental signature of their formation in fusion reactions. One of the earliest attempts to study cluster radioactivity in the superheavy region was done by Poenaru et al. [21]. In this work nuclear life-times for cluster radioactivity of superheavy elements and nuclei far off the β-stability line for $Z = 52$--$122$ have been calculated. A simple theoretical approach was proposed by Singh et al. [22] to study cluster decay of some heavy and super heavy nuclei [23, 24]. The proposed model is based on the fragmentation theory [25] and on the preformed cluster decay model proposed by Gupta and Malik [26]. Based on the concept of cold valley, Kumar et al. [27] studied cluster emission from superheavy nucleus $^{277}$Ds and from its α-decay products $^{273}$Hs and $^{269}$Sg. They concluded that in addition to α-decay and fission $^{14}$C, $^{34}$Si and $^{50}$Ca are optimal cases of cluster decay. However the predicted half-lives are of a huge order of magnitude larger than the expected compound nuclei life-times. Recently Poenaru et al. [28] studied heavy particle radioactivity, the emission of clusters with $Z > 28$ from superheavy nuclei with $Z = 104$--124.

In the background of all these studies this paper presents a study of cluster radioactivity from the $^{286}$112, $^{292}$114 and $^{296}$116 superheavy nuclei using the Coulomb and Proximity Potential Model (CPPM) [29] proposed by one of us (KPS) in 2000. The proximity potential has been extensively used by one of us (KPS) in cluster radioactivity studies [30, 31] and in heavy-ion-induced fusion [32]. The excited compound nuclei $^{286}$112, $^{292}$114 and $^{296}$116 have been obtained during the fusion processes with $^{48}$Ca beam on $^{238}$U, $^{244}$Pu and $^{248}$Cm at the same excitation energy $E^*$ = 33 MeV [33], and could lead to the formation of superheavy nuclei in $3n$--$4n$ evaporation channels. In the post-scission region the interacting barrier is taken as the sum of Coulomb and Proximity potential and for the overlap (pre-scission) region simple power law interpolation [34] is used.

## 2. THE COULOMB AND PROXIMITY POTENTIAL MODEL

The interacting potential barrier for a parent nucleus exhibiting cluster decay is given by

$$V = \frac{Z_1 Z_2 e^2}{r} + V_p(z) + \frac{\hbar^2 \ell(\ell+1)}{2\mu r^2}, \text{ for } z > 0. \tag{1}$$

Here, $Z_1$ and $Z_2$ are the atomic numbers of daughter and emitted cluster, '$r$' is the distance between fragment centers, $\ell$ the angular momentum and $V$p is the proximity potential given by Blocki et al. [35]. The reduced mass, $\mu = mA_1 A_2 / A$ , where m is the nucleon mass and $A_1$, $A_2$ are the mass numbers of daughter and emitted cluster respectively.

$$V_p(z) = 4\pi\gamma b\left[\frac{C_1 C_2}{(C_1 + C_2)}\right]\Phi\left(\frac{z}{b}\right), \tag{2}$$

with the nuclear surface tension coefficient

$$\gamma = 0.9517[1 - 1.7826(N - Z)^2 / A^2] \qquad [\text{MeV/fm}^2]. \tag{3}$$

$N$, $Z$, and $A$ represent neutron, proton, and mass numbers of parent and $\Phi$ represents the universal proximity potential [36] given by

$$\Phi(\varepsilon) = -4.41 e^{-\varepsilon / 0.7176} \quad \text{, for } \varepsilon \geq 1.9475, \tag{4}$$

$$\Phi(\varepsilon) = -1.7817 + 0.9270\varepsilon + 0.0169\varepsilon^2 - 0.05148\varepsilon^3 \quad \text{for } 0 \leq \varepsilon \leq 1.9475, \tag{5}$$

with $\varepsilon = z/b$, where the width (diffuseness) of the nuclear surface $b \approx 1$ and Sussmann central radius $C_i$ of fragments related to sharp radius $R_i$ is:

$$C_i = R_i - \left(\frac{b^2}{R_i}\right). \tag{6}$$

For $R_i$ we use semi-empirical formula in terms of mass number $A_i$ as

$$R_i = 1.28 A_i^{1/3} - 0.76 + 0.8 A_i^{-1/3} . \tag{7}$$

Using one-dimensional WKB approximation, the barrier penetrability $P$ is got as

$$P = \exp\left\{-\frac{2}{\hbar}\int_{\varepsilon_0}^{\varepsilon_f}\sqrt{2\mu(V - Q)}\,dz\right\}. \tag{8}$$

Here, the mass parameter is replaced by the reduced mass. Here, $\varepsilon_0 = 2(C - C_1 - C_2)$ and $\varepsilon_f$ is defined as $V(\varepsilon_f) = Q$ . The decay energy $Q$ is given by,

$$Q = M(A, Z) - M(A_1, Z_1) - M(A_2, Z_2)\,, \tag{9}$$

where $M(A, Z)$, $M(A_1, Z_1)$, $M(A_2, Z_2)$ are the atomic masses of parent, daughter and emitted cluster respectively. The above integral can be evaluated numerically or analytically, and the half-life time is given by

$$T_{1/2} = \left(\frac{\ln 2}{\lambda}\right) = \left(\frac{\ln 2}{\upsilon P}\right), \tag{10}$$

where $\upsilon = \left(\frac{\omega}{2\pi}\right) = \left(\frac{2E_v}{h}\right)$ represents the number of assaults on the barrier per second and $\lambda$ is the decay constant. $E_v$, the empirical zero point vibration energy [37] is given by

$$E_v = Q\left\{0.056 + 0.039\exp\left[\frac{(4 - A_2)}{2.5}\right]\right\} \qquad \text{for } A_2 \geq 4. \tag{11}$$

## 3. RESULTS AND DISCUSSION

The study of possible cluster decay has been done using CPPM and it reveals that there is a possibility for cluster emission from the selected parents with half-life times well within the present upper limit for measurements $(T_{1/2} < 10^{30}\text{s})$. The study is based on the concept of cold valley which was introduced in relation to the structure of minima in the so-called driving potential. The driving potential is defined as the difference between the interaction potential and the decay energy $Q$ of the reaction. Most of the $Q$ values are calculated using experimental mass excesses of Audi et al. [38]. The interaction potential is calculated in two steps – first it is taken as just the Coulomb potential, while in the second step it is calculated as the sum of Coulomb and Proximity Potentials. Next the driving potential $(V - Q)$ for a particular parent is calculated for all possible fragments as a function of mass and charge asymmetries, $\eta = \frac{A_1 - A_2}{A_1 + A_2}$ and $\eta_Z = \frac{Z_1 - Z_2}{Z_1 + Z_2}$, at the touching configuration. For every fixed mass pair ($A_1$, $A_2$) a pair of charges is singled out for which driving potential is minimum. Figures 1--3 represent the plots for driving potential versus $A_2$ (mass of one fragment) for $^{286}112$, $^{292}114$ and $^{296}116$ parents respectively. There are two curves for each parent – one calculated without the proximity potential and another one including the proximity potential. The occurrence of the mass – asymmetry valleys in these figures is due to the shell effects. It is found that inclusion of the proximity potential does not change the position of the minima but they become deeper.

For $^{286}112$ in addition to the alpha particle $^{8}$Be, $^{10}$Be, $^{14}$C, $^{24}$Ne, $^{28}$Mg, $^{34}$Si, $^{50}$Ca, etc. are found to be possible candidates for emission. Moving on to the fission region, there are two deep regions each consisting of three comparable minima. For the first valley, as one can see from Fig.1, the first minimum corresponds to the splitting $^{78}$Zn + $^{208}$Pb, while the second and third minima correspond to the splittings $^{80}$Ge + $^{206}$Hg and $^{82}$Ge + $^{204}$Hg. From the cold valley approach the first minimum is due to the double magicity of $^{208}$Pb, the second minimum is occurring due to the magic neutron shell $N$ = 126 of $^{206}$Hg, while the third splitting is occurring due to the magic neutron shell $N$ = 50 of $^{82}$Ge. In the case of the second valley, the first two minima involve $^{130}$Sn + $^{156}$Sm and $^{132}$Sn + $^{154}$Sm splittings and therefore their occurrence is attributed to the presence of $Z$ = 50 and $Z$ = 50, $N$ = 82 magic shells respectively. The third minimum comes from the splitting $^{134}$Te + $^{152}$Nd and is due to the magic neutron shell $N$ = 82 of $^{134}$Te.

Just as in the case of $^{286}112$, $^{10}$Be, $^{14}$C, $^{24}$Ne, $^{28}$Mg, $^{34}$Si, $^{40}$S, $^{50}$Ca, etc. are detected to be possible clusters for $^{292}114$ and $^{296}116$ and are displayed in Figs 2 and 3 respectively. Similarly, in the fission region two deep regions with three minima each are observed. For $^{292}114$ the first valley corresponds to the splittings $^{82}$Ge + $^{210}$Pb, $^{84}$Ge + $^{208}$Pb and $^{86}$Se + $^{206}$Hg. Here the first minimum can be explained in terms of magic shells $N = 50$ and $Z = 82$ of $^{82}$Ge and $^{210}$Pb respectively. The second minimum is of course due to the doubly-magic $^{208}$Pb nucleus while the third minimum is due to the magic shell $N = 126$ of $^{206}$Hg. In the case of the second deep region of $^{292}114$ plot the minima are due to the splittings $^{130}$Sn + $^{162}$Gd, $^{132}$Sn + $^{160}$Gd and $^{134}$Te + $^{158}$Sm. The first two are attributed to the magicity of $^{130}$Sn and $^{132}$Sn nuclei, while the third one is due to the magic shell $N = 82$ of $^{134}$Te. Finally considering the $^{296}116$ nucleus, the first deep region has the splittings $^{86}$Se + $^{210}$Pb, $^{88}$Se + $^{208}$Pb and $^{90}$Kr + $^{206}$Hg, while the second one gives the splittings $^{130}$Sn + $^{166}$Dy, $^{132}$Sn + $^{164}$Dy and $^{134}$Te + $^{162}$Gd. Just as in the case of $^{292}114$, all these minima can be explained in terms of magic shells of Pb, Hg, Sn and Te nuclei.

The half-life times and other characteristics for alpha decay and clusters up to $^{68}$Ni have been calculated for all the three parents and presented in Tables 1--3. The computed alpha half-lives are in good agreement with the values calculated using Viola--Seaborg semi-empirical relationship with constants determined by Sobiczewski, Patyk and Cwiok [39] which is given by

$$\log_{10} T_{1/2} = (aZ + b)/\sqrt{Q} + cZ + d \,, \qquad (12)$$

where the half-life is in seconds, $Q$ value is in MeV, and $Z$ is the atomic number of the parent nucleus. Instead of using the original set of constants determined by Viola and Seaborg [40] more recent values $a = 1.66175$, $b = -8.5166$, $c = -0.20228$ and $d = -33.9069$ that were determined in an adjustment taking into account the new data for even-even nuclei are used. The $T_{1/2}$ values are also calculated using the Scaling Law of Horoi et al. [41] for cluster decay and compared with the present values. The Scaling Law is given by the equation,

$$\log_{10} T_{1/2} = (a_1\mu^x + b_1)[(Z_1Z_2)^y / \sqrt{Q} - 7] + (a_2\mu^x + b_2)\,, \qquad (13)$$

where $\mu$ is the reduced mass. The six parameters are $a_1$ = 9.1, $b_1$ = -10.2, $a_2$ = 7.39, $b_2$ = -23.2, $x$ = 0.416 and $y$ = 0.613.

On examining the $T_{1/2}$ values one finds that in addition to the α-particle $^{8}$Be and $^{14}$C clusters are also probable for emission from the three chosen nuclei $^{286}$112, $^{292}$114 and $^{296}$116 with $T_{1/2} \leq 10^{30}$s which is well within the present upper limit for measurements. In Figures 1--3 the valley left to $^{10}$Be cluster is that of $^{8}$Be cluster, so both clusters are probable to appear for emission from the respective parent nucleus. From Table 1--3 it is clear that $T_{1/2}$ value for the $^{8}$Be cluster emission from $^{286}$112, $^{292}$114, $^{296}$116 are 4.56 x $10^{30}$s, 1.69 x $10^{29}$s, 1.69 x $10^{21}$s respectively, which lie near and within the measurable upper limit. The computed half-life value for $^{10}$Be cluster emission is found to be $T_{1/2} > 10^{38}$s far away from experimental upper limit so that the $^{8}$Be is the optimum nucleus for cluster emission.

Figures 4--6 represent the plot connecting computed logarithm of half-life versus mass number of emitted cluster from $^{286}$112, $^{292}$114, $^{296}$116 parents respectively and their comparison with Viola--Seaborg--Sobiczewski systematics (VSS) and Scaling Law of Horoi et al., [41]. It is found that our prediction agrees with VSS and Scaling Law. We have compared cluster decay half-life with the results reported by Poenaru et al. [21, 28] based on ASAFM and it is found that our values agree well with them. For example in case of $^4$He emission from $^{296}$116 isotope $\log_{10}(T_{1/2}^{\mathrm{CPPM}}) = -2.58$, $\log_{10}(T_{1/2}^{\mathrm{ASAFM}}) = -3.20$; in the case of $^8$Be emission $\log_{10}(T_{1/2}^{\mathrm{CPPM}}) = 21.23$, $\log_{10}(T_{1/2}^{\mathrm{ASAFM}}) = 21.04$.

We have computed half-lives for various clusters ranging from $^4$He to $^{70}$Ni from superheavy $^{280\text{-}314}$116 isotopes [42] and found that $^4$He emission from $^{280}$116 and $^{302}$116 parents have minimum half-life compared to other clusters. This indicates the role of doubly-magic ${}^{276}_{162}114$ and ${}^{298}_{184}114$ daughter nuclei in the corresponding decays. The study on cluster decay from $^{294\text{–}326}$122 [31] also reveal the presence of doubly-magic ${}^{304}_{184}120$ and ${}^{298}_{184}114$ daughter. Both the study and the reported high stability of $^{285}$114 against α decay [13] reveal that the cluster radioactivity in the region of superheavy nuclei is controlled by shell effects in parent (daughter) nucleus like in the case of heavy nuclei, where cluster decay leads to a magic daughter nucleus around $^{208}$Pb.

The possible alpha-decay chains from $^{286}$112, $^{292}$114 and $^{296}$116 are also studied and the corresponding characteristics are given in Table 4. The computed alpha-decay half-lives are compared with the values predicted by Generalized Liquid Drop Model (GLDM) of Royer [43] and they are found to match reasonably well.

## 4. CONCLUSIONS

Using the Coulomb and Proximity Potential model proposed by one of us (KPS) in 2000, theoretical estimates for cluster emission from superheavy elements $^{286}$112, $^{292}$114 and $^{296}$116 are presented. It is found that in addition to alpha particle $^8$Be and $^{14}$C clusters are also most probable for emission from these nuclei. Although none of these cases of cluster emission is recorded till date, this study is expected to give an understanding of the behavior of these newly synthesized nuclei during possible decay. Again considering the fission of these nuclei all the cases of possible splittings could be explained in terms of magicity of one or both fragments. The calculated alpha half-lives are in good agreement with those calculated using VSS systematics. The possible alpha decay chains of the three parent nuclei are tabulated and wherever available the corresponding alpha half-lives are compared with those calculated using GLDM systematics.

One of the authors (S. S.) would like to thank the Kannur University, Kerala, India for financial support in the form of Junior Research Fellowship.

REFERENCES

1. A. Sandulescu, D. N. Poenaru, W. Greiner, Fiz. Elem. Chastits At. Yadra **11**, 1334 (1980) [Sov. J. Part. Nucl. **11**, 528 (1980)].
2. R. K. Gupta, Sovt. J. Part. Nucl. **8**, 289 (1978).
3. J. A. Maruhn, W. Greiner, W. Scheid, *Heavy Ion Collisions* (North-Holland, Amsterdam, 1980), Vol. 2, p. 399.
4. H. J. Rose and G. A. Jones, Nature **307**, 245 (1984).
5. D. N. Poenaru, D. Schnabel, W. Greiner, *et al.*, At. Data. Nucl. Data. Tables **48**, 231 (1991).
6. Yu. Ts. Oganessian, V. L. Mikheev, S. P. Tretyakova, JINR Report No. E 7 - 93 - 57 (Dubna, 1993).
7. Yu. Ts. Oganessian, Yu. A. Lazarev, V. L. Mikheev, *et al.*, Z. Phys. A **349**, 341 (1994).
8. A. Guglielmetti, B. Blank, R. Bonetti, *et al.*, Nucl. Phys. A **583,** 867c (1995).
9. A. Guglielmetti, R. Bonetti, G. Poli, *et al.*, Phys. Rev. C **52**, 740 (1995).
10. A. Sobiczewski, F. A. Gareev, B. N. Kalinkin, Phys. Lett. **22**, 500 (1966).
11. H. Meldner, Ark. Fys. **36**, 593 (1967).
12. S. G. Nilsson, C. F. Tsang, A. Sobiczewski, *et al.*, Nucl. Phys. A **131**, 1 (1969).
13. U. Mosel and W. Greiner, Z. Phys. A **222**, 261 (1969).

14. S. Cwiok, J. Dobaczewski, P. H. Heenen, *et al*., Nucl. Phys. A **611**, 211 (1996).
15. K. Rutz, M. Bender, T. Burvenich, *et al*., Phys. Rev. C **56**, 238 (1997).
16. S. K. Patra, W. Greiner, R. K. Gupta, J. Phys. G **26**, L65 (2000).
17. S. B. Duarte, O. A. P. Tavares, M. Goncalves, *et al*., J. Phys. G **30**, 1491 (2004).
18. B. K. Sharma, P. Arumugam, S. K. Patra, *et al*., J. Phys. G **32**, L9 (2006).
19. R. K. Gupta, Sushil Kumar, Rajesh Kumar, *et al*., J. Phys. G **28,** 2879 (2002).
20. Yu. V. Denisov and S. Hofmann, Phys. Rev. C **61,** 034606 (1999).
21. D. N. Poenaru, D. Schnabel, W. Greiner, *et al*., At. Data. Nucl. Data Tables **48,** 231 (1991).
22. S. Singh, R. K. Gupta, W. Scheid, W. Greiner, J. Phys. G **18**, 1243 (1992).
23. R. K. Gupta, S. Singh, R. K. Puri, W. Scheid, Phys. Rev. C **47**, 561 (1992).
24. R. K. Gupta, S. Singh, G. Münzenberg, W. Scheid, Phys. Rev. C **51**, 2623 (1995).
25. H. J. Fink, J. A. Marhun, W. Scheid, W. Greiner, Z. Phys. A **268,** 321 (1974).
26. S. S. Malik, S. Singh, R. K. Puri, *et al*., Pramana - J. Phys. **32**, 419 (1989).
27. S. Kumar, M. Balasubramaniam, R. K. Gupta, *et al*., J. Phys. G **29**, 625 (2003).
28. D. N. Poenaru, R. A. Gherghescu, W. Greiner, arXiv: 1106.3271v1 [nucl-th].
29. K. P. Santhosh and A. Joseph, Pramana - J. Phys **55**, 375 (2000).
30. K. P. Santhosh, R. K. Biju, A. Joseph, J. Phys. G. **35**, 085102 (2008).
31. K. P. Santhosh and R. K. Biju, J. Phys. G **36**, 015107 (2009).
32. K. P. Santhosh, V. Bobby Jose, A. Joseph, K. M. Varier, Nucl. Phys. A **817**, 35 (2009).
33. M. G. Itkis et al., in Proceedings of the International Workshop on Fusion Dynamics at the Extremes, Dubna, 2000, Ed. by Yu. Ts. Oganessian and V. I. Zagrebaev (World Sci., Singapore, 2001) p. 93.
34. Y. J. Shi and W. J. Swiatecki, Nucl. Phys. A **438**, 450 (1985).
35. J. Blocki, J. Randrup, W. J. Swiatecki, C. F. Tsang, Ann. Phys (N.Y.) **105**, 427 (1977).
36. J. Blocki and W. J. Swiatecki, Ann. Phys (N.Y.) **132**, 53 (1981).

37. D. N. Poenaru, M. Ivascu, A. Sandulescu, W. Greiner, Phys. Rev. C **32**, 572 (1985).

38. G. Audi, A. H. Wapstra, C. Thibault, Nucl. Phys. A **729,** 337 (2003).

39. A. Sobiczewski, Z. Patyk, S. Cwiok, Phys. Lett. B **224**, 1 (1989).

40. V. E. Viola and G. T. Seaborg, J. Inorg. Nucl. Chem. **28**, 741 (1966).

41. M. Horoi, A. Brown and A. Sandulescu, arXiv: 9403008v1 [nucl-th].

42. K. P. Santhosh and R. K. Biju, Pramana - J. Phys **72**, 689 (2009).

43. G. Royer, J. Phys. G **26,** 1149 (2000).

**Table 1**. Computed half-life time values and other characteristics for the cluster decay of $^{286}112$ nucleus

| Parent nucleus | Emitted cluster | Daughter nuclei | $Q$ value, MeV | Penetrability $P$ | Decay constant $\lambda$, $s^{-1}$ | $\log_{10}(T_{1/2})$ |
|---|---|---|---|---|---|---|
| $^{286}112$ | $^{4}$He | $^{282}110$ | 8.986 | $4.4566 \times 10^{-25}$ | $1.8399 \times 10^{-4}$ | 3.58 |
| | $^{8}$Be | $^{278}$Hs | 17.199 | $2.8591 \times 10^{-52}$ | $1.5190 \times 10^{-31}$ | 30.66 |
| | $^{10}$Be | $^{276}$Hs | 13.874 | $5.1029 \times 10^{-72}$ | $2.0385 \times 10^{-51}$ | 50.53 |
| | $^{14}$C | $^{272}$Sg | 35.131 | $1.4834 \times 10^{-52}$ | $1.4294 \times 10^{-31}$ | 30.69 |
| | $^{20}$O | $^{266}$Rf | 49.794 | $1.2214 \times 10^{-62}$ | $1.6490 \times 10^{-41}$ | 40.62 |
| | $^{24}$Ne | $^{262}$No | 68.448 | $5.0104 \times 10^{-63}$ | $9.2900 \times 10^{-42}$ | 40.87 |
| | $^{26}$Ne | $^{260}$No | 66.660 | $4.9079 \times 10^{-67}$ | $8.8611 \times 10^{-46}$ | 44.89 |
| | $^{28}$Mg | $^{258}$Fm | 87.298 | $9.3239 \times 10^{-64}$ | $2.2045 \times 10^{-42}$ | 41.50 |
| | $^{32}$Mg | $^{254}$Fm | 82.602 | $1.0533 \times 10^{-72}$ | $2.3564 \times 10^{-51}$ | 50.47 |
| | $^{34}$Si | $^{252}$Cf | 106.629 | $5.0905 \times 10^{-63}$ | $1.4700 \times 10^{-41}$ | 40.67 |
| | $^{38}$S | $^{248}$Cm | 122.175 | $1.5868 \times 10^{-68}$ | $5.2505 \times 10^{-47}$ | 46.12 |
| | $^{40}$S | $^{246}$Cm | 122.938 | $2.0805 \times 10^{-66}$ | $6.9267 \times 10^{-45}$ | 44.00 |
| | $^{42}$S | $^{244}$Cm | 121.453 | $2.7683 \times 10^{-68}$ | $9.1054 \times 10^{-4}$ | 45.88 |
| | $^{44}$Ar | $^{242}$Pu | 140.249 | $9.2882 \times 10^{-68}$ | $3.5279 \times 10^{-46}$ | 45.29 |
| | $^{46}$Ar | $^{240}$Pu | 142.299 | $5.5553 \times 10^{-63}$ | $2.1409 \times 10^{-41}$ | 40.51 |
| | $^{48}$Ca | $^{238}$U | 159.612 | $7.6676 \times 10^{-65}$ | $3.3144 \times 10^{-43}$ | 42.32 |
| | $^{50}$Ca | $^{236}$U | 159.831 | $4.6465 \times 10^{-63}$ | $2.0113 \times 10^{-41}$ | 40.54 |
| | $^{52}$Ca | $^{234}$U | 157.060 | $1.0621 \times 10^{-66}$ | $4.5178 \times 10^{-45}$ | 44.19 |
| | $^{54}$Ti | $^{232}$Th | 173.017 | $1.7276 \times 10^{-69}$ | $8.0949 \times 10^{-48}$ | 46.93 |
| | $^{56}$Ti | $^{230}$Th | 170.943 | $8.3283 \times 10^{-72}$ | $3.8556 \times 10^{-50}$ | 49.25 |
| | $^{58}$Cr | $^{228}$Ra | 185.694 | $1.1669 \times 10^{-75}$ | $5.8685 \times 10^{-54}$ | 53.07 |
| | $^{60}$Cr | $^{226}$Ra | 185.838 | $6.7438 \times 10^{-74}$ | $3.3941 \times 10^{-52}$ | 51.31 |
| | $^{62}$Cr | $^{224}$Ra | 185.082 | $8.3023 \times 10^{-74}$ | $4.1615 \times 10^{-52}$ | 51.22 |
| | $^{64}$Fe | $^{222}$Rn | 201.433 | $1.7370 \times 10^{-73}$ | $9.4757 \times 10^{-52}$ | 50.86 |
| | $^{66}$Fe | $^{220}$Rn | 202.396 | $6.0025 \times 10^{-70}$ | $3.2902 \times 10^{-48}$ | 47.32 |
| | $^{68}$Ni | $^{218}$Po | 217.834 | $3.5959 \times 10^{-70}$ | $2.1214 \times 10^{-48}$ | 47.51 |

**Table 2**. The same as Table 1 but for the $^{292}114$ nucleus

| Parent nucleus | Emitted cluster | Daughter nuclei | $Q$ value, MeV | Penetrability $P$ | Decay constant $\lambda$, $s^{-1}$ | $\log_{10}(T_{1/2})$ |
|---|---|---|---|---|---|---|
| $^{292}114$ | $^{4}He$ | $^{288}112$ | 9.626 | $1.1722 \times 10^{-23}$ | $5.1838 \times 10^{-3}$ | 2.13 |
| | $^{8}Be$ | $^{284}110$ | 18.019 | $7.3570 \times 10^{-51}$ | $4.0950 \times 10^{-30}$ | 29.23 |
| | $^{10}Be$ | $^{282}110$ | 14.524 | $1.9546 \times 10^{-70}$ | $8.1740 \times 10^{-50}$ | 48.93 |
| | $^{14}C$ | $^{278}Hs$ | 34.841 | $2.7734 \times 10^{-55}$ | $2.6503 \times 10^{-34}$ | 33.42 |
| | $^{20}O$ | $^{272}Sg$ | 50.074 | $1.2128 \times 10^{-64}$ | $1.6466 \times 10^{-43}$ | 42.62 |
| | $^{22}O$ | $^{270}Sg$ | 49.270 | $1.5772 \times 10^{-67}$ | $2.1057 \times 10^{-46}$ | 45.52 |
| | $^{24}Ne$ | $^{268}Rf$ | 70.768 | $1.5623 \times 10^{-61}$ | $2.9949 \times 10^{-40}$ | 39.36 |
| | $^{26}Ne$ | $^{266}Rf$ | 68.880 | $1.4154 \times 10^{-65}$ | $2.6405 \times 10^{-44}$ | 43.42 |
| | $^{28}Mg$ | $^{264}No$ | 89.658 | $4.4420 \times 10^{-63}$ | $1.0786 \times 10^{-41}$ | 40.81 |
| | $^{30}Mg$ | $^{262}No$ | 87.840 | $2.9764 \times 10^{-66}$ | $7.0807 \times 10^{-45}$ | 43.99 |
| | $^{32}Si$ | $^{260}Fm$ | 107.470 | $2.9414 \times 10^{-66}$ | $8.5611 \times 10^{-45}$ | 43.91 |
| | $^{34}Si$ | $^{258}Fm$ | 107.957 | $6.4753 \times 10^{-65}$ | $1.8932 \times 10^{-43}$ | 42.56 |
| | $^{36}Si$ | $^{256}Fm$ | 105.340 | $4.0543 \times 10^{-69}$ | $1.1566 \times 10^{-47}$ | 46.78 |
| | $^{38}S$ | $^{254}Cf$ | 123.946 | $2.4117 \times 10^{-70}$ | $8.0952 \times 10^{-49}$ | 47.93 |
| | $^{40}S$ | $^{252}Cf$ | 125.242 | $2.5586 \times 10^{-67}$ | $8.6782 \times 10^{-46}$ | 44.90 |
| | $^{42}S$ | $^{250}Cf$ | 124.454 | $6.7371 \times 10^{-68}$ | $2.2707 \times 10^{-46}$ | 45.48 |
| | $^{44}Ar$ | $^{248}Cm$ | 143.296 | $3.8384 \times 10^{-68}$ | $1.4896 \times 10^{-46}$ | 45.67 |
| | $^{46}Ar$ | $^{246}Cm$ | 145.528 | $5.6451 \times 10^{-63}$ | $2.2249 \times 10^{-41}$ | 40.49 |
| | $^{48}Ar$ | $^{244}Cm$ | 143.173 | $3.8418 \times 10^{-66}$ | $1.4896 \times 10^{-44}$ | 43.67 |
| | $^{50}Ca$ | $^{242}Pu$ | 163.278 | $1.4984 \times 10^{-63}$ | $6.6259 \times 10^{-42}$ | 41.02 |
| | $^{52}Ca$ | $^{240}Pu$ | 160.799 | $1.3572 \times 10^{-66}$ | $5.9104 \times 10^{-45}$ | 44.07 |
| | $^{54}Ti$ | $^{238}U$ | 176.877 | $4.5227 \times 10^{-70}$ | $2.1665 \times 10^{-48}$ | 47.51 |
| | $^{56}Ti$ | $^{236}U$ | 175.080 | $8.1459 \times 10^{-72}$ | $3.8624 \times 10^{-50}$ | 49.25 |
| | $^{58}Cr$ | $^{234}Th$ | 189.741 | $8.9463 \times 10^{-77}$ | $4.5972 \times 10^{-55}$ | 54.18 |
| | $^{60}Cr$ | $^{232}Th$ | 189.777 | $3.7060 \times 10^{-75}$ | $1.9047 \times 10^{-53}$ | 52.56 |
| | $^{62}Cr$ | $^{230}Th$ | 188.763 | $1.7332 \times 10^{-75}$ | $8.8604 \times 10^{-54}$ | 52.89 |
| | $^{64}Fe$ | $^{228}Ra$ | 204.584 | $3.7437 \times 10^{-77}$ | $2.0742 \times 10^{-55}$ | 54.52 |
| | $^{66}Fe$ | $^{226}Ra$ | 205.058 | $1.4989 \times 10^{-74}$ | $8.3240 \times 10^{-53}$ | 51.92 |
| | $^{68}Ni$ | $^{224}Ra$ | 219.506 | $9.1702 \times 10^{-78}$ | $5.4514 \times 10^{-56}$ | 55.10 |

**Table 3**. The same as Table 1 but for the $^{296}116$ nucleus

| Parent nucleus | Emitted cluster | Daughter nuclei | $Q$ value, MeV | Penetrability $P$ | Decay constant $\lambda$, $s^{-1}$ | $\log_{10}(T_{1/2})$ |
|---|---|---|---|---|---|---|
| $^{296}116$ | $^{4}He$ | $^{292}114$ | 11.506 | $4.9953 \times 10^{-19}$ | $2.6406 \times 10^{2}$ | -2.58 |
| | $^{8}Be$ | $^{288}112$ | 21.039 | $6.2913 \times 10^{-43}$ | $4.0887 \times 10^{-22}$ | 21.23 |
| | $^{10}Be$ | $^{286}112$ | 17.044 | $2.6616 \times 10^{-60}$ | $1.3062 \times 10^{-39}$ | 38.72 |
| | $^{14}C$ | $^{282}110$ | 38.041 | $1.3540 \times 10^{-49}$ | $1.4127 \times 10^{-28}$ | 27.69 |
| | $^{20}O$ | $^{276}Hs$ | 52.334 | $3.3542 \times 10^{-62}$ | $4.7595 \times 10^{-41}$ | 40.16 |
| | $^{22}O$ | $^{274}Hs$ | 51.140 | $8.2133 \times 10^{-66}$ | $1.1381 \times 10^{-44}$ | 43.78 |
| | $^{24}Ne$ | $^{272}Sg$ | 73.748 | $5.1262 \times 10^{-59}$ | $1.0241 \times 10^{-37}$ | 36.83 |
| | $^{26}Ne$ | $^{270}Sg$ | 72.050 | $2.0636 \times 10^{-62}$ | $4.0271 \times 10^{-41}$ | 40.24 |
| | $^{28}Mg$ | $^{268}Rf$ | 93.768 | $1.3819 \times 10^{-59}$ | $3.5094 \times 10^{-38}$ | 37.30 |
| | $^{30}Mg$ | $^{266}Rf$ | 92.120 | $3.6082 \times 10^{-62}$ | $9.0020 \times 10^{-41}$ | 39.87 |
| | $^{32}Si$ | $^{264}No$ | 112.650 | $6.8712 \times 10^{-62}$ | $2.0963 \times 10^{-40}$ | 39.52 |
| | $^{34}Si$ | $^{262}No$ | 112.107 | $3.8784 \times 10^{-62}$ | $1.1775 \times 10^{-40}$ | 39.77 |
| | $^{36}Si$ | $^{260}No$ | 109.140 | $9.6847 \times 10^{-67}$ | $2.8626 \times 10^{-45}$ | 44.38 |
| | $^{38}S$ | $^{258}Fm$ | 128.791 | $3.3237 \times 10^{-67}$ | $1.1593 \times 10^{-45}$ | 44.78 |
| | $^{40}S$ | $^{256}Fm$ | 129.720 | $1.0201 \times 10^{-64}$ | $3.5835 \times 10^{-43}$ | 42.29 |
| | $^{42}S$ | $^{254}Fm$ | 128.652 | $1.1573 \times 10^{-65}$ | $4.0323 \times 10^{-44}$ | 43.24 |
| | $^{44}Ar$ | $^{252}Cf$ | 148.584 | $6.1349 \times 10^{-65}$ | $2.4687 \times 10^{-43}$ | 42.45 |
| | $^{46}Ar$ | $^{250}Cf$ | 150.904 | $6.1056 \times 10^{-60}$ | $2.4952 \times 10^{-38}$ | 37.44 |
| | $^{48}Ca$ | $^{248}Cm$ | 169.179 | $9.1189 \times 10^{-62}$ | $4.1780 \times 10^{-40}$ | 39.22 |
| | $^{50}Ca$ | $^{246}Cm$ | 169.309 | $6.3476 \times 10^{-60}$ | $2.9105 \times 10^{-38}$ | 37.38 |
| | $^{52}Ca$ | $^{244}Cm$ | 166.403 | $1.4654 \times 10^{-63}$ | $6.6041 \times 10^{-42}$ | 41.02 |
| | $^{54}Ti$ | $^{242}Pu$ | 183.397 | $1.8297 \times 10^{-66}$ | $9.0878 \times 10^{-45}$ | 43.88 |
| | $^{56}Ti$ | $^{240}Pu$ | 181.329 | $1.6160 \times 10^{-68}$ | $7.9356 \times 10^{-47}$ | 45.94 |
| | $^{58}Cr$ | $^{238}U$ | 196.977 | $7.4345 \times 10^{-73}$ | $3.9660 \times 10^{-51}$ | 50.24 |
| | $^{60}Cr$ | $^{236}U$ | 196.710 | $1.4047 \times 10^{-71}$ | $7.4835 \times 10^{-50}$ | 48.97 |
| | $^{62}Cr$ | $^{234}U$ | 195.410 | $2.9315 \times 10^{-72}$ | $1.5514 \times 10^{-50}$ | 49.65 |
| | $^{64}Fe$ | $^{232}Th$ | 212.007 | $1.0620 \times 10^{-73}$ | $6.0976 \times 10^{-52}$ | 51.06 |
| | $^{66}Fe$ | $^{230}Th$ | 211.793 | $3.6083 \times 10^{-72}$ | $2.0697 \times 10^{-50}$ | 49.52 |
| | $^{68}Ni$ | $^{228}Ra$ | 226.900 | $1.9721 \times 10^{-75}$ | $1.2119 \times 10^{-53}$ | 52.76 |

**Table 4**. Computed half-life time values and other characteristics for the alpha decay chains of $^{286}112$, $^{292}114$ and $^{296}116$ nuclei and their comparison with GLDM

| Parent nucleus | Daughter | $Q$ value, MeV | Penetrability P | Decay constant $\lambda$, $s^{-1}$ | $\log_{10}(T_{1/2})$ | |
|---|---|---|---|---|---|---|
| | | | | | present | GLDM |
| $^{286}112$ | $^{282}110$ | 8.986 | $4.4566 \times 10^{-25}$ | $1.8399 \times 10^{-4}$ | 3.58 | 2.84 |
| $^{282}110$ | $^{278}Hs$ | 8.306 | $8.8377 \times 10^{-27}$ | $3.3725 \times 10^{-6}$ | 5.31 | 4.81 |
| $^{278}Hs$ | $^{274}Sg$ | 8.476 | $2.4336 \times 10^{-25}$ | $9.4768 \times 10^{-5}$ | 3.86 | |
| $^{274}Sg$ | $^{270}Rf$ | 8.236 | $1.9837 \times 10^{-25}$ | $7.5061 \times 10^{-5}$ | 3.97 | |
| $^{270}Rf$ | $^{266}No$ | 7.476 | $1.1416 \times 10^{-27}$ | $3.9211 \times 10^{-7}$ | 6.25 | |
| $^{296}116$ | $^{292}114$ | 11.506 | $4.9953 \times 10^{-19}$ | $2.6406 \times 10^{2}$ | -2.58 | -3.67 |
| $^{292}114$ | $^{288}112$ | 9.626 | $1.1722 \times 10^{-23}$ | $5.1838 \times 10^{-3}$ | 2.13 | |
| $^{288}112$ | $^{284}110$ | 8.486 | $7.2757 \times 10^{-27}$ | $2.8366 \times 10^{-6}$ | 5.39 | |
| $^{284}110$ | $^{280}Hs$ | 8.386 | $1.9143 \times 10^{-26}$ | $7.3755 \times 10^{-6}$ | 4.97 | |
| $^{280}Hs$ | $^{276}Sg$ | 7.686 | $2.0034 \times 10^{-28}$ | $7.0743 \times 10^{-8}$ | 6.99 | |

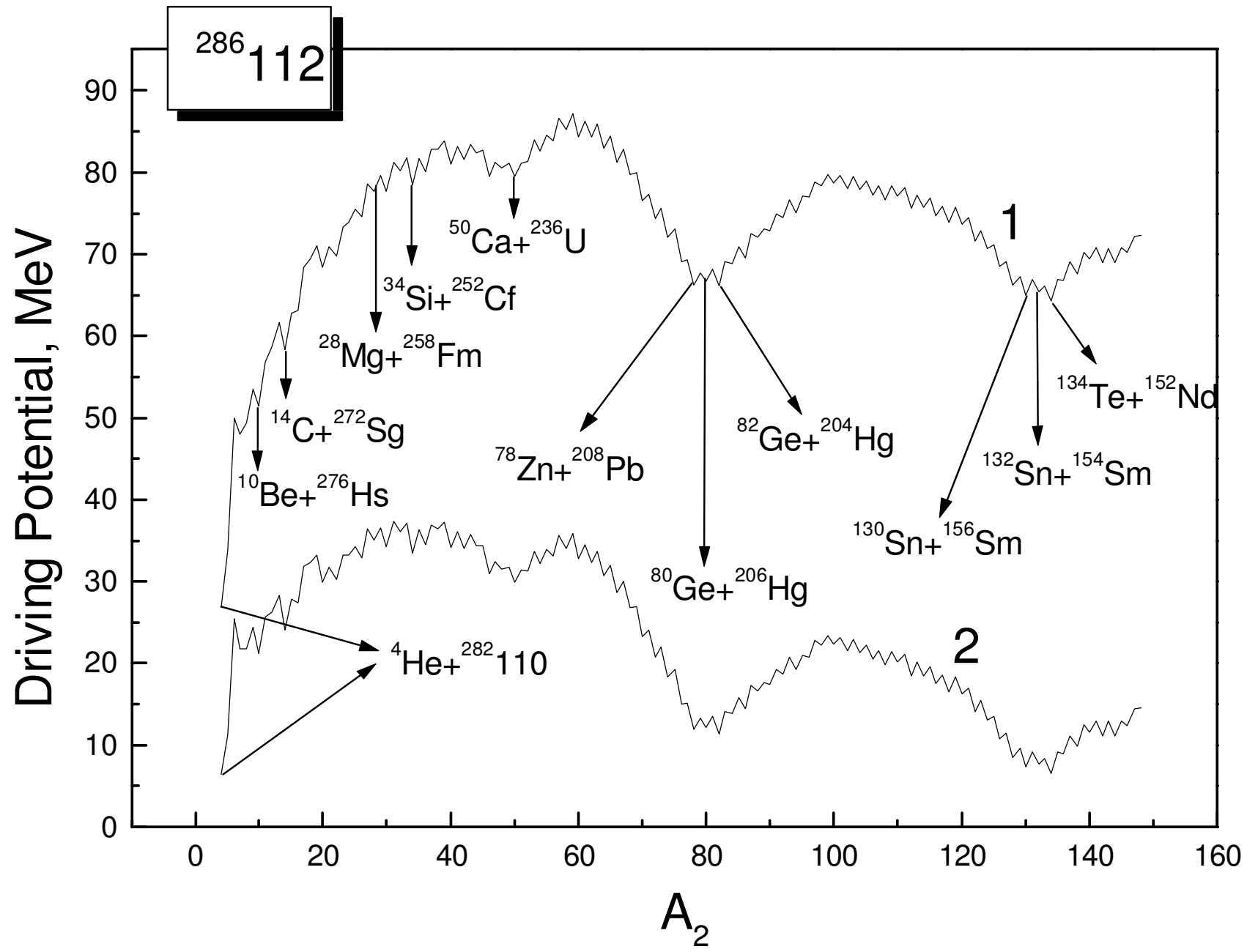

**Fig. 1**. Plot for driving potential as a function of mass of one fragment $A_2$ calculated at the touching configuration (1) without proximity potential, (2) with proximity potential for $^{286}112$ nucleus.

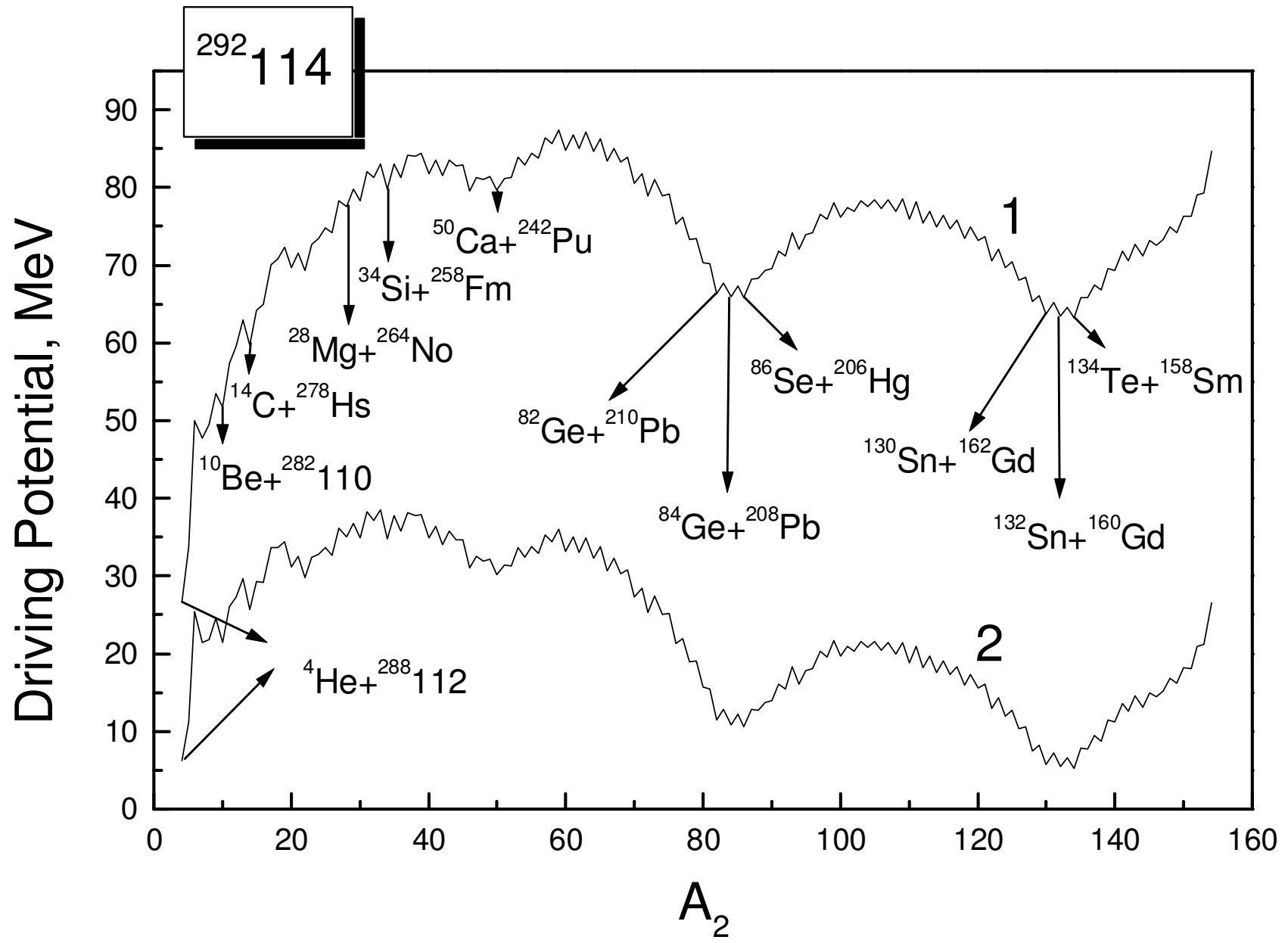

**Fig. 2**. The same as in Fig. 1, but for $^{292}114$ nucleus.

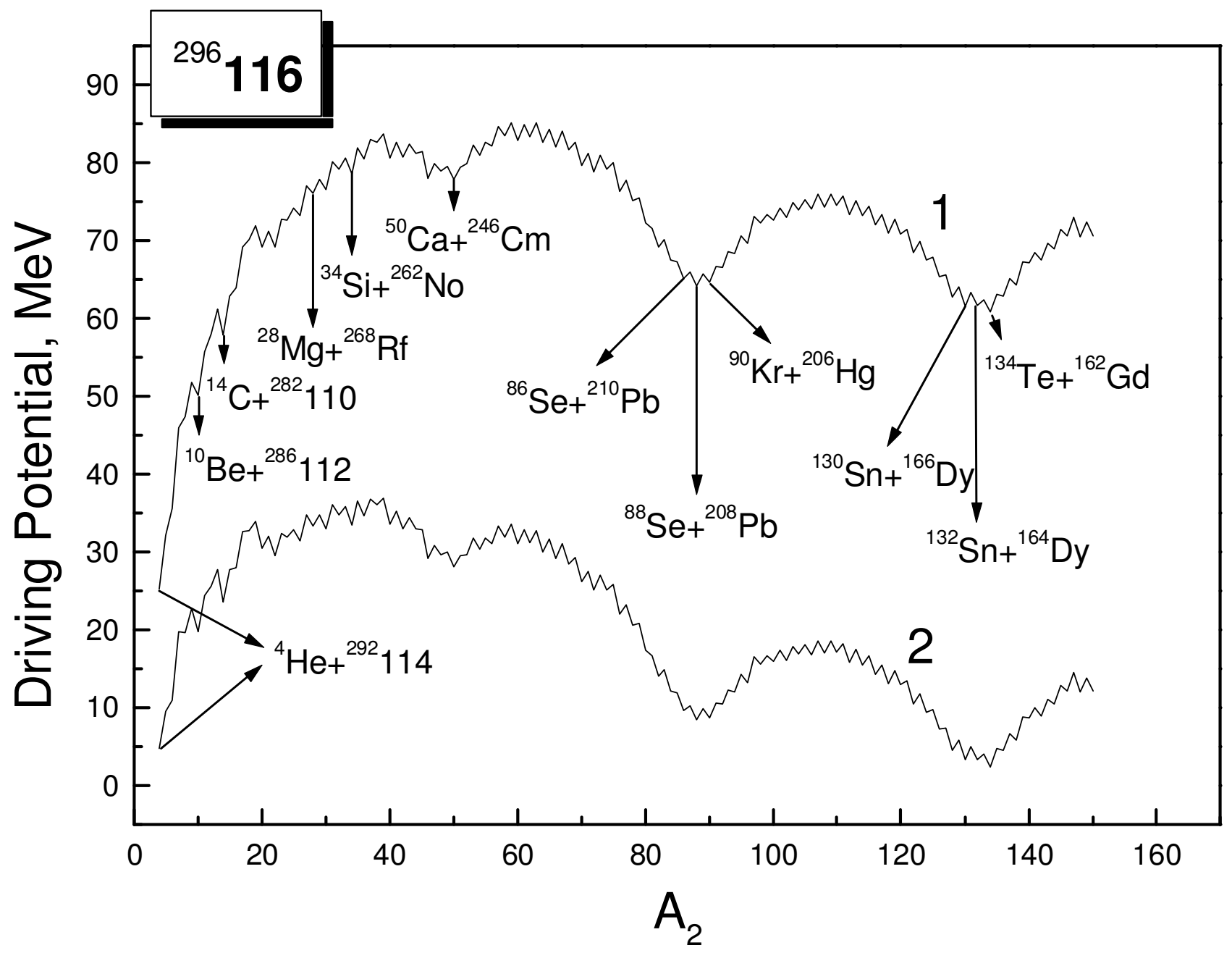

**Fig. 3.** The same as in Fig. 1, but for $^{296}$116 nucleus.

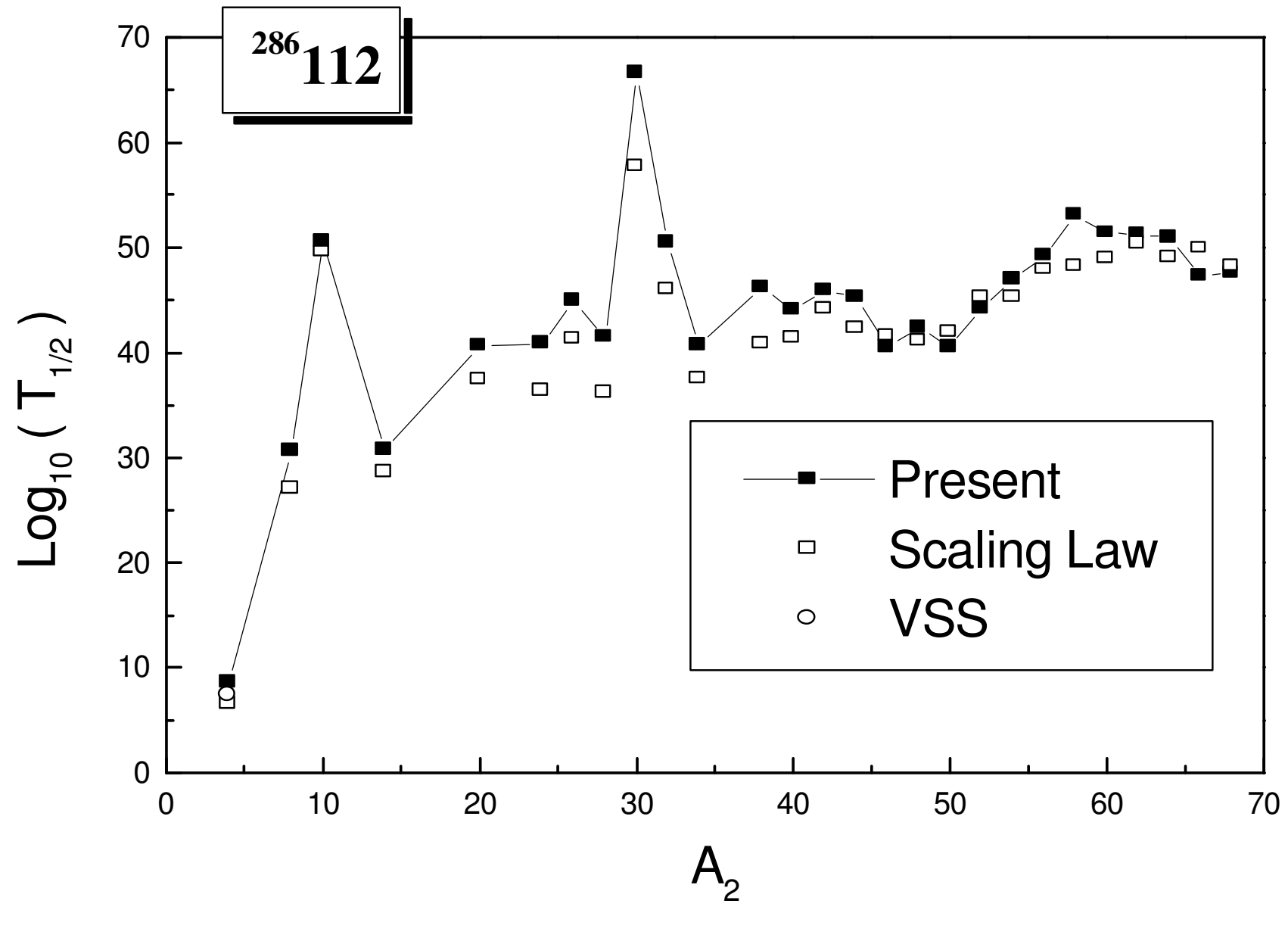

**Fig. 4.** Plot for logarithm of half life time vs mass number of cluster for $^{286}$112 parent.

.

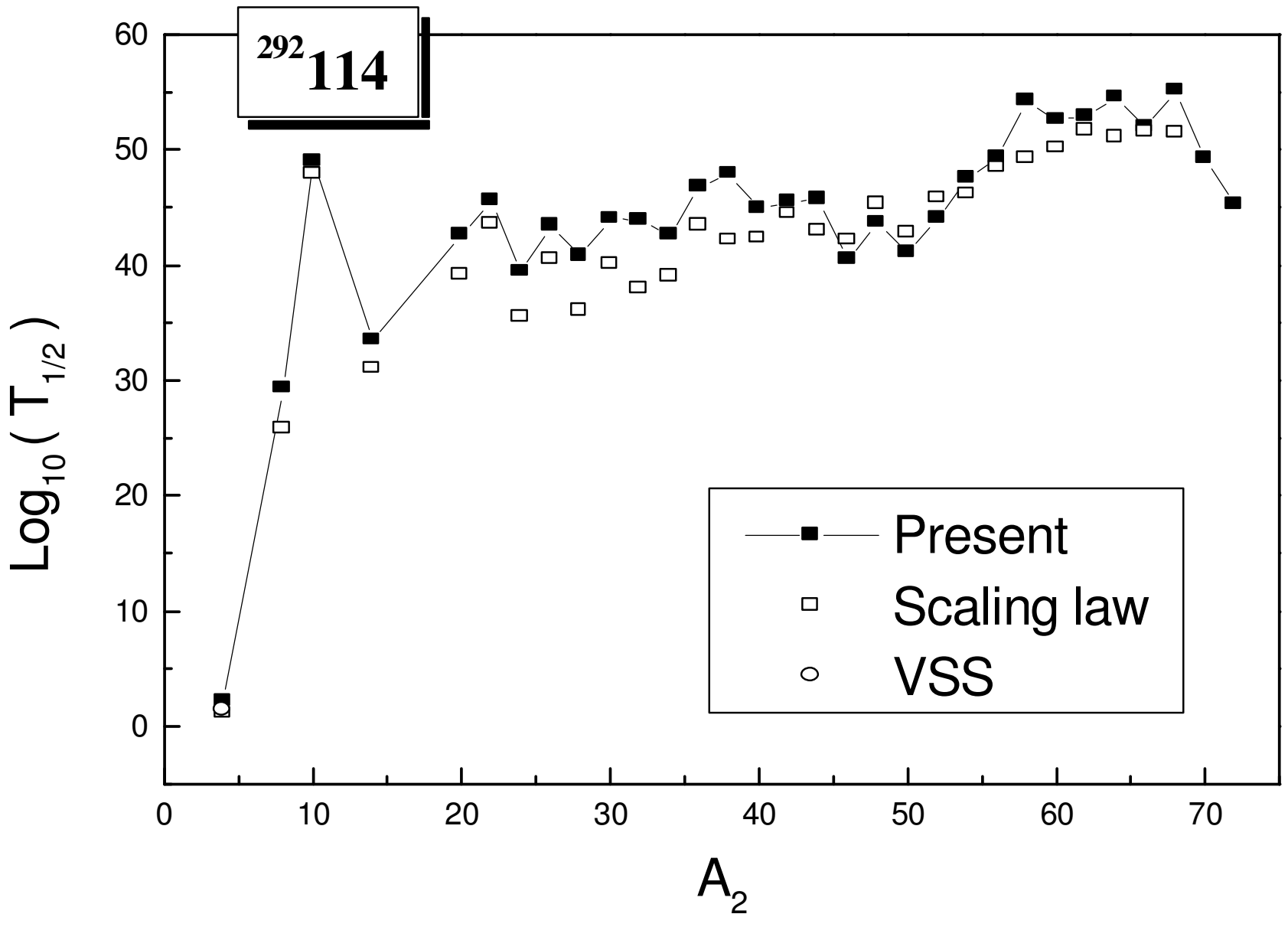

**Fig. 5.** The same as in Fig. 4, but for $^{292}114$ parent.

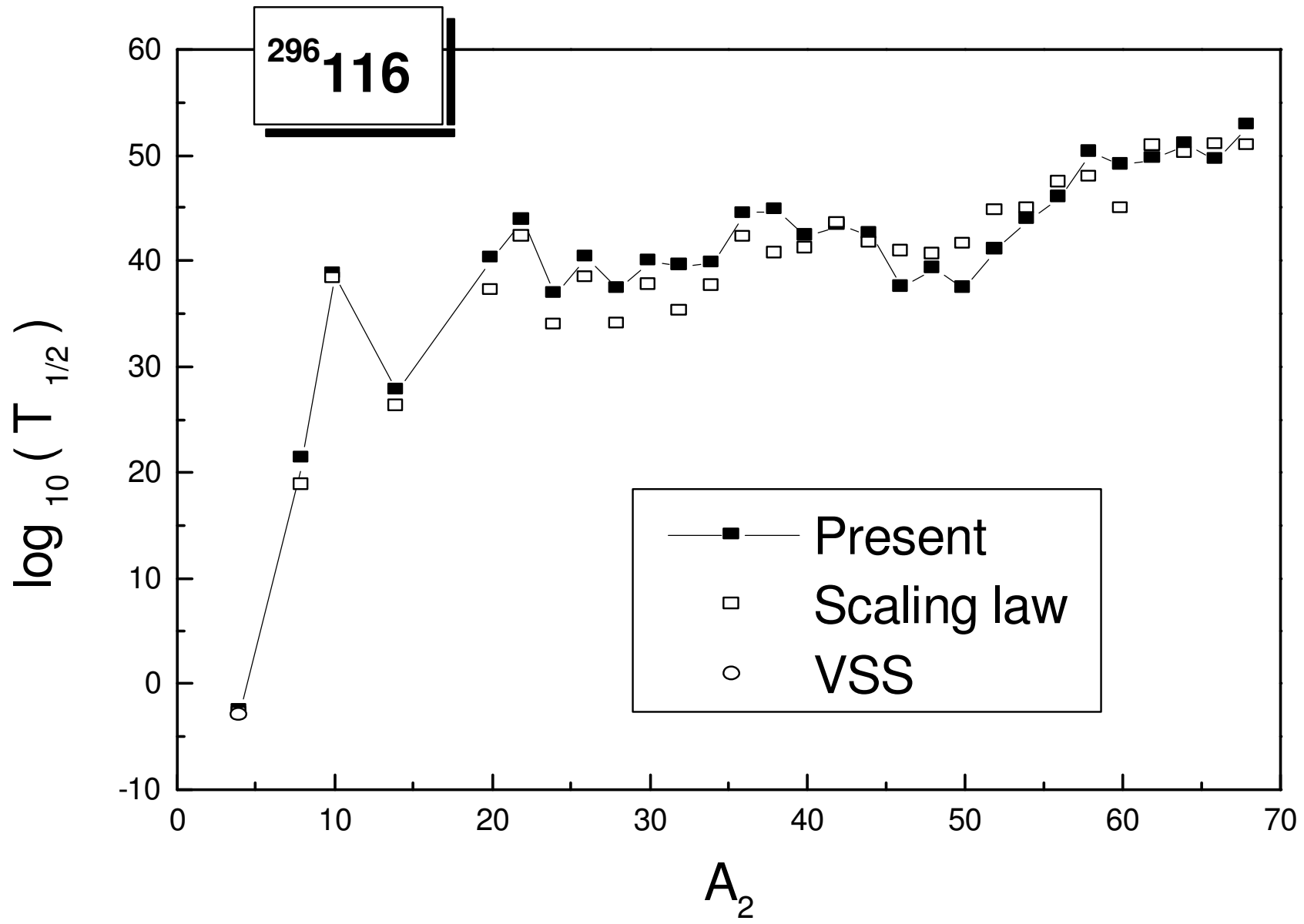

**Fig. 6.** The same as in Fig. 4, but for $^{296}116$ parent.